\documentclass[10pt,conference,letterpaper]{IEEEtran}
\IEEEoverridecommandlockouts
\usepackage{cite}
\usepackage{amsthm,amsmath,amssymb,amsfonts}
\usepackage{graphicx}
\usepackage{textcomp}
\usepackage{xcolor}
\usepackage{algorithm}
\usepackage{gensymb}
\usepackage{algpseudocode}
\usepackage{subcaption}
\usepackage{geometry}
\theoremstyle{definition}

\newtheorem{Theorem}{Theorem}

\newtheorem{Problem}{Problem}
\newtheorem{Note}[Theorem]{Note}

\newcommand{\R}{\mathbb{R}}

\newcommand{\V}{\mathcal{V}}
\newcommand{\im}{\mbox{Im}}
\newcommand{\krnl}{\mbox{Ker}}

\def\BibTeX{{\rm B\kern-.05em{\sc i\kern-.025em b}\kern-.08em
    T\kern-.1667em\lower.7ex\hbox{E}\kern-.125emX}}
\begin{document}

\title{Disturbance Decoupling Problem for $n$-link chain pendulum on a cart system }

\author{\IEEEauthorblockN{Sayar Das}

\and
\IEEEauthorblockN{Deepak U. Patil}

\thanks{}%
\thanks{The authors are with the Department of Electrical Engineering, Indian Institute of Technology Delhi, Hauz Khas, New Delhi, 110016, India (email: sayar29897@gmail.com; deepakpatil@ee.iitd.ac.in)}}
\maketitle
\begin{abstract}
    A disturbance decoupling problem for a $n$-link chain pendulum on a cart is considered. A model of the cart developed in a coordinate-free framework and the linearized equations of this system are considered from \cite{b1}. It is shown that it is possible to design a suitable state feedback such that the angular position or velocity of the $n^{th}$-link can always be decoupled from the disturbance coming at the cart.
\end{abstract}
\section{Introduction}
 Classical pendulum models have provided a wealth of examples, challenges, and applications in nonlinear dynamics and control. As such, many problems of linear and non-linear control have been solved for various pendulum models in the literature of control systems. One such important problem is the problem of disturbance decoupling. The disturbance decoupling problem seeks to choose a state feedback such that the disturbance input to a system does not affect the output or a variable of interest of the system. As such, this problem finds significance in many applications like robotic arms, LIGO, designing PCBs and system neuroscience.

 Of particular interest among the various pendulum models found in the literature is the $n$-link chain pendulum on a cart. The $n$-link pendulum on a cart system is a classical and important control system model that has been worked on extensively in previous literature. One of the papers that covers this topic extensively and will be helpful for this paper is \cite{b1}, where the authors develop the Euler-Lagrange equation for the system in a coordinate-free framework as well as provide a linearization about each equilibrium point. We will take the linear model of the system developed by the authors in the aforementioned paper. The disturbance decoupling problem can then be solved for the linear model. Variations of this problem have been solved in several ways in previous literature. In \cite{b9}, using the approach of geometric control, the solution to the disturbance decoupling problem has been obtained through PID control laws. In \cite{b7}, the disturbance decoupling issue has been resolved for a robotic manipulator on a moving platform by feedback linearization. The disturbance decoupling problem was also examined for a robotic manipulator where the mechanical dynamics of the links and electrical dynamics of the joint motors were included in the equations of motion \cite{b8}. In this paper, the authors, using feedforward control canceled the disturbance signal in the signal flow graph of the closed-loop control system, thus eliminating the effects of the disturbance signal. 

 As mentioned earlier, one of the applications of the disturbance decoupling problem applied to a pendulum model is the highly sensitive measurement systems used in Laser Interferometer Gravitational Objects (LIGO) which are used to measure gravitational waves \cite{b4}. In the LIGO, there is a quadruple pendulum and the test mass is attached at the end of the quadruple pendulum. The disturbance decoupling problem in LIGO attempts to decouple the displacement of the test mass from the ground disturbance.
\begin{figure}
  \centering
  \includegraphics[width=1\linewidth]{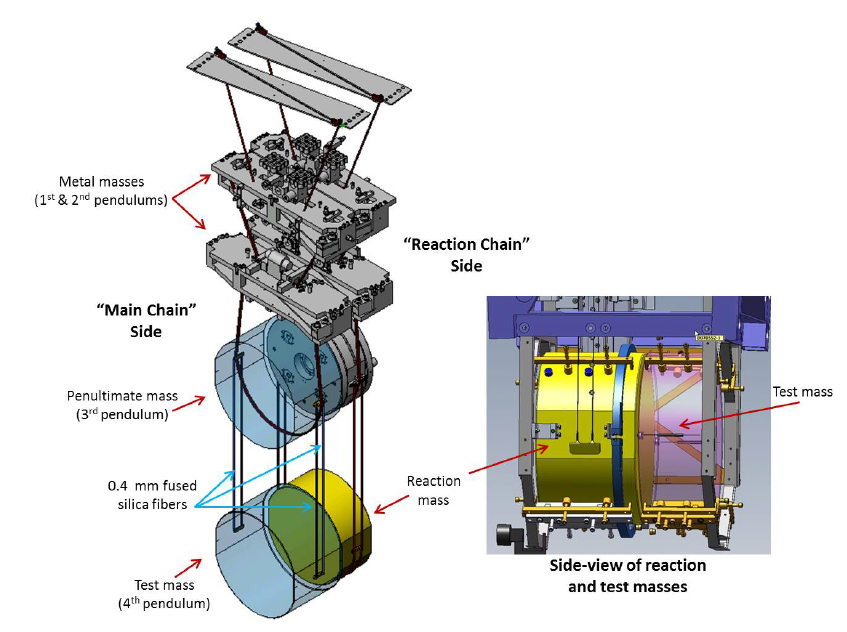}
  \label{fig:sfig1}
\end{figure}%
\begin{figure}
  \centering
  \includegraphics[width=1\linewidth]{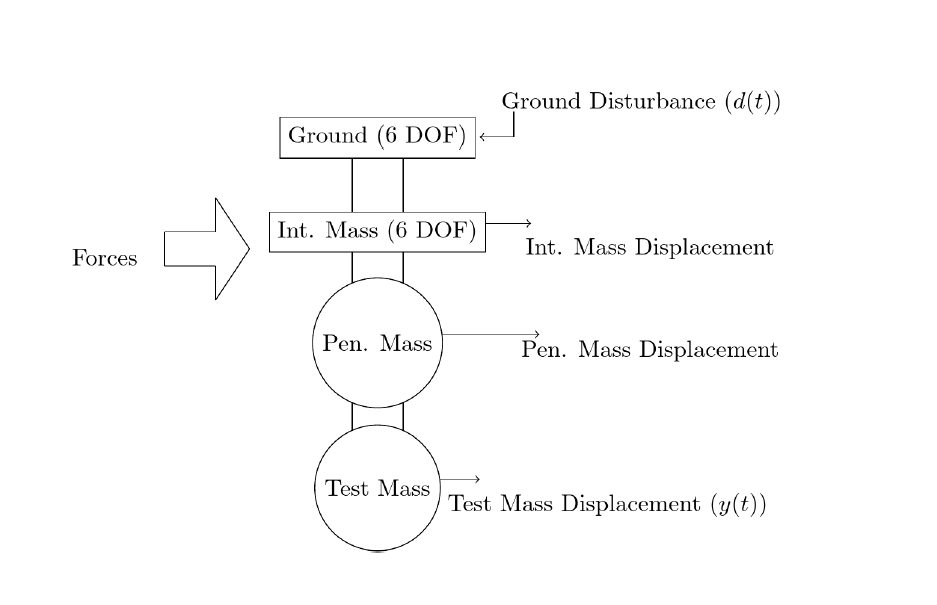}
  \caption{Vibration Isolation in LIGO}
  \label{fig:sfig2}
\end{figure}

 The problem of disturbance decoupling applied to LIGO motivates us to solve the problem for a generalized $n$-link chain pendulum on a cart system in this paper.

 In this paper, we consider the disturbance decoupling problem for a general $n$-link pendulum on a cart, where we assume the disturbance comes at the cart. We establish that state and prove a theorem which states that it is possible to design a state feedback such that the position of the $n^{th}$-link of the pendulum can be decoupled from the disturbance at the cart, provided $n \geq 2$. We do this by proving that the condition of successful disturbance decoupling holds for this system.

 The rest of the paper is organized as follows. Section 2 reviews the theory of disturbance decoupling and some known results and formulates the problem. In Section 3, we take the model of $n$-link chain pendulum on a cart and the equations of motion for the system from \cite{b1} and linearize the system with reference to the same paper. In Section 4, we propose a theorem that it is possible to design a state feedback such that the position of the $n^{th}$-link of the pendulum can be decoupled from the disturbance at the cart, provided $n \geq 2$. Section 5 presents some simulations of some cases of the system with and without feedback when it is attacked by a disturbance input. This is followed by the conclusion in Section 6.
\section{Problem Formulation}
 Consider a linear time-invariant system be described by the following equation:
\begin{equation}\label{system_equation}
    \begin{aligned}
        \dot{x} &= Ax + Bu + Ew\\
        y &= Hx
    \end{aligned}
\end{equation}
where $A \in \R^{n\times n}$ is the system matrix, $B \in \R^{n \times m}$ is the input matrix, $E \in \R^{n \times d}$ is the disturbance input matrix and $H \in \R^{h \times n}$ is the output matrix. A goal of disturbance decoupling problem is to compute $F$ s.t. setting the state feedback as $u = Fx$, decouples the disturbance from the output. The solution to equation  \eqref{system_equation} is given by : 
\begin{equation}\label{system_equation_solution}y(t) = He^{(A+BF)t}x(0) + \int_{0}^{t} He^{(A+BF)(t-\tau)}Ew(\tau)\, d\tau \end{equation}
Now if we want that the disturbance does not affect $y(t)$, then we need
\begin{align}\label{dist_decoup}
H\begin{bmatrix}
E &  (A+BF)E & .& . & . & (A+BF)^{n-1}E
 \end{bmatrix} &= 0 
\end{align}
Define the subspace $\mathcal{V}^*$ as follows:
 \begin{equation*}
 \mathcal{V}^* := \im(\begin{bmatrix}
 E &  (A+BF)E & . & . &. & (A+BF)^{n-1}E
 \end{bmatrix})
\end{equation*}
It is shown in \cite{b2} that the subspace  $\mathcal{V}^*$ is the smallest $(A+BF)$ - invariant  subspace which contains image of $E$ and is contained inside kernel of $H$. Further, any matrix $F$ s.t. $(A+BF)\mathcal{V}^*\subset \mathcal{V}^*$ is called as a Friend of $\mathcal{V}^*$. Now, the output $y$ is decoupled from the disturbance $w$ if and only if there exists a $(A+BF)$-invariant subspace, $\mathcal{V}^*$ which contains image of $E$ and is contained inside Kernel of $H$ (see \cite{b2} and \cite{b3}).
\begin{equation}\label{dist_decoup_cond}
 \begin{aligned}
    \im(E)\subset\mathcal{V}^*\subset \krnl(H)
    \end{aligned}
\end{equation}

Thus, if we can find the $\mathcal{V}^*$ and an associated feedback matrix $F$ i.e., a Friend of $\mathcal{V}^*$  s.t. \eqref{dist_decoup_cond} holds, then by setting $u=Fx$, we achieve perfect disturbance decoupling. 

The procedure for finding the controlled invariant subspace $\V^*$ and the feedback matrix $F$ can be found in detail from \cite{b2},\cite{b3}.

Some properties of controlled invariant subspaces will be crucial in the following sections, hence it is stated below in the form of a theorem \cite{b2},\cite{b3}.
\begin{Theorem}\label{properties}
Let $\mathcal{V}$ be a controlled invariant subspace. Then the following statements are equivalent:

\begin{enumerate}
    \item $\mathcal{V}$ is a controlled invariant subspace
    \item $A\mathcal{V} \subset \mathcal{V} + \mbox{Im}(B)$
    \item $\exists F: \mathbb{R}^n \rightarrow \mathbb{R}$ such that $(A+BF)\mathcal{V}\subset\mathcal{V}$ 
\end{enumerate}
\end{Theorem}

The system of a $n$-link chain pendulum on a cart can be analyzed to see if a controlled invariant subspace $\mathcal{V}^*$ can be found which will achieve perfect disturbance decoupling of the $n^{th}$-link from the disturbance coming at the cart. 

\subsection{Problem Formulation}
 The problem statement can be formally stated as,
 \begin{Problem}\label{Problem2}
     Given the linearized system of an $n$-link pendulum on a cart system about an equilibrium point, prove that it is possible to design a state feedback such that the position of the $n^{th}$-link of the pendulum can be decoupled from the disturbance coming at the cart, provided $n \geq 2$.
 \end{Problem}

The Problem \ref{Problem2} is resolved if there exists a state feedback matrix $F$ such that condition (\ref{dist_decoup_cond}) is satisfied. To this end, we take the model of the $n$-link pendulum on a cart system developed in a coordinate-free framework from \cite{b1} with slight modifications to the state vector. Then we use the structure of the $A$, $B$, $E$ and $H$ matrices to prove that it is always possible to construct an $(A,B)$-invariant subspace for this system. Hence this system always meets the disturbance decoupling condition (\ref{dist_decoup_cond}).

\section{Model of $n$-link chain pendulum on a cart}
\begin{figure}[h!]
    \centering
    \includegraphics[scale = 0.55]{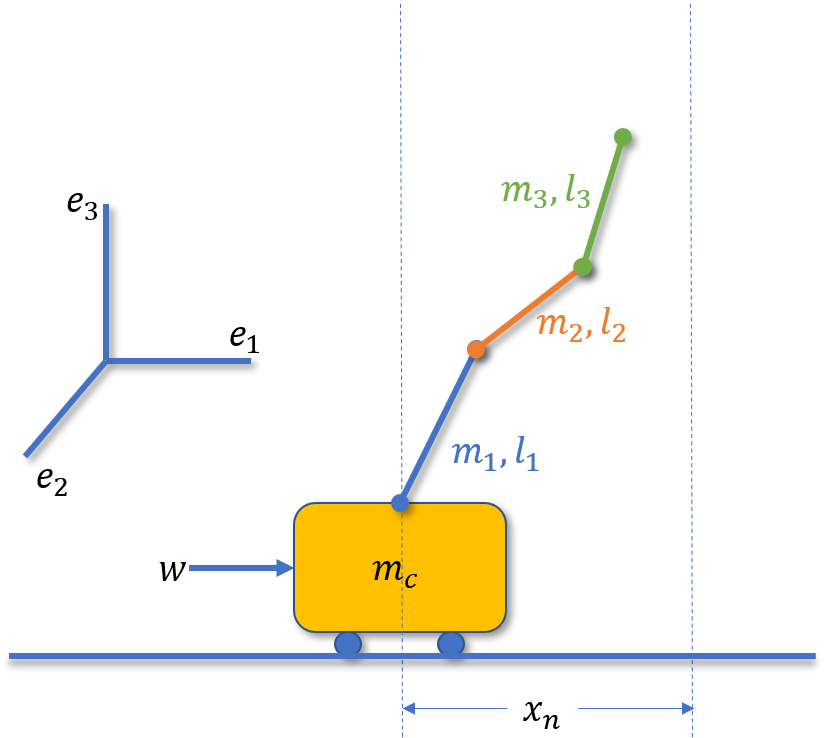}
    \caption{$n$-link chain pendulum on a cart}
\end{figure}
The model, equations and linearized equations of the $n$-link pendulum on a cart is taken from \cite{b1}. Slight modifications have been made to the state vector to suit the purpose of the theorem we will be proving in Section 4. The model and relevant equations are restated here for the sake of clarity and convenience.

The $n$-link pendulum is attached to a cart that moves in a horizontal plane and is made up of $n$ rigid links that are connected by a spherical joint \cite{b1}. In the equations which will follow, $x \in \mathbb{R}^2$ denotes the displacement of the cart, $v \in \R^2$ is the velocity of the cart, $q_i \in \R^3$ are the direction vectors of each link in an inertial frame and $\omega_i \in \R^3$ are the angular velocities of the links in the same inertial frame \cite{b1}. The manifold $S^2$ is defined as follows:
\begin{equation*}
    \{q \in \mathbb{R}^3 |\hspace{0.1cm} \|q\|_2 = 1\}
\end{equation*}
Hence, the configuration manifold of this system is given by $(S^2)^n \times \mathbb{R}^2$ because taking unit vectors as the direction vectors of the $n$ links, the $n$ links move in a unit sphere and the cart moves in a horizontal plane.

The inertial frame, as mentioned before is given by the usual Cartesian coordinates
i.e.
\begin{equation*}\begin{aligned}e_1 =\begin{bmatrix}
1\\
0\\
0\\
\end{bmatrix}; e_2 =\begin{bmatrix}
0\\
1\\
0\\
\end{bmatrix}; e_3 =\begin{bmatrix}
0\\
0\\
1\\
\end{bmatrix}\end{aligned}.\end{equation*} Here $e_3$ is considered along the direction of gravity. Matrix $C$ is defined as:
\begin{equation*}
    C = \left[e_1 \middle | e_2 \right]
\end{equation*}

The direction vectors $q_i$ are related to the angular velocity $\omega_i$ by the following kinematic equation:
\begin{equation}\label{kinematic_eq}
    \begin{aligned}
        \dot{q_i} = \omega_i \times q_i = \hat{\omega}_iq_i
    \end{aligned}
\end{equation}
where $\hat{\omega_i}$ denotes a map $\R^3 \rightarrow \mathfrak{so}(3)$ takes a vector in $\R ^3$ and provides a skew-symmetric matrix that is only specified by the following equation,
\begin{equation}\label{skew}
\begin{aligned}
    \hat{x}y = x \times y = \begin{bmatrix}
        0 & x_2 & -x_3\\
        -x_1 & 0 & x_3\\
        x_1 & -x_2 & 0
    \end{bmatrix}\begin{bmatrix}
        y_1\\
        y_2\\
        y_3
    \end{bmatrix}
    \end{aligned}
\end{equation}
where \begin{equation*}
\begin{aligned}
    x &= \begin{bmatrix}
        x_1 & x_2 & x_3
    \end{bmatrix}^\intercal\\
    y &= \begin{bmatrix}
        y_1 & y_2 & y_3
    \end{bmatrix}^\intercal
    \end{aligned}
\end{equation*}
\subsection{Euler-Lagrange equations}
The Euler-Lagrangian equation on $(S^2)^n \times \R^2$ in terms of the angular velocities, $\omega_i$ is given as follows:
\begin{equation}\label{EL_eqn_for_n_link_pendulum}
    \begin{aligned}
    &\begin{bmatrix}
        M_{00} & M_{01}\hat{q_1} & . & . & . & M_{0n}\hat{q_n}\\
 -\hat{q_1} M_{10} & M_{11} I_3 & . &. &. & -M_{1n}\hat{q_1}\hat{q_n}\\
 .\\
 .\\
 -\hat{q_n} M_{n0} & -M_{n1} \hat{q_n}\hat{q_1} &. &. &. & M_{nn} I_3
 \end{bmatrix} \begin{bmatrix}
 \Ddot{x}\\
 \dot{\omega_1}\\
 .\\
 .\\
 \dot{\omega_n} 
 \end{bmatrix} \\ &= \begin{bmatrix}
 \sum_{j=1}^{n}M_{0j}\|\omega_j\|^2q_j + u\\
 \sum_{j=2}^{n}M_{1j}\|\omega_j\|^2\hat{q_1}q_j + \sum_ {a = 1}^n m_a g l_1 \hat{q_1} e_3\\
 \sum_{j=1,j \neq 2}^{n}M_{1j}\|\omega_j\|^2\hat{q_2}q_j + \sum_ {a = 2}^n m_a g l_1 \hat{q_1} e_3\\
 .\\
 .\\
 \sum_{j=1}^{n-1}M_{nj}\|\omega_j\|^2\hat{q_n}q_j + m_n g l_n \hat{q_n} e_3\\
 \end{bmatrix}\\
 \dot{q_i} &= \omega_i \times q_i
    \end{aligned}    
\end{equation}
where the blocks are defined as follows:
\begin{equation*}
    \begin{aligned}
        M_{00} &= m + \sum_ {i = 1}^n m_i\\
        M_{0i} &= C^\intercal \sum_ {a = i}^n m_a l_i\\
        M_{i0} &= M_{0i}^\intercal\\
        M_{ij} &= \left(\sum_{a = \max\{i,j\}}^{n}m_a\right)l_il_j
    \end{aligned}
\end{equation*}
for $i,j = 1,2,...,n$.
\subsection{Equilibrium Configurations}
The equilibrium configurations of this $n$-link pendulum system are those configurations where all the links are directed along or opposite to the direction of gravity. i.e.
\begin{equation}\label{equilibrium}
 q_i \times e_3 = 0
 \end{equation}
 
To better handle the different equilibrium situations \cite{b1}, we introduce a variable $s$ which is defined as :
\begin{equation}\label{equilibrium_variable}
\begin{aligned}
s = 1  &\big{|} q_i = e_3\\
s = -1 &\big{|} q_i = -e_3
\end{aligned}
\end{equation}

An ordered n-tuple $\{s_1,s_2, \ldots, s_n \}$ will be taken as a particular equilibrium configuration.
\subsection{Linearized equations of motion}
The variation of a curve $q_i(t)$ on $S^2$ can be written as:
\begin{equation}
    q_i^\epsilon(t) = \mbox{exp}(\epsilon \xi(t))q_i(t)
\end{equation}
where $\xi(t) \in \R^3$ such that $q_i(t).\xi(t) = 0$ for all time, $t$. The exponential term belongs to $SO(3)$, the set of all orthogonal matrices which form the 3-D rotation group \cite{b1}.

Hence the corresponding infinitesimal variation \cite{b1} is:
\begin{equation}\label{variation_of_q}
    \delta q_i(t) = \xi(t) \times q_i(t)
\end{equation}
which lies in the tangential space $T_{q_{i}(t)}S^2 $. 

From (\ref{equilibrium}), we see the equilibrium points are when the links point in the direction of the $e_3$- axis. Hence variation of $q_i$ from any equilibrium configuration can be written as:
\begin{equation}\label{variation_from_equilibrium}
    \delta q_i = \xi \times e_3
\end{equation}
where $\xi .e_3 = 0$. The variation of $\omega \in \R^3$ is given by $\delta \omega \in \R^3$ where $\delta \omega . e_3 = 0$. For any equilibrium configuration, there is no component of $\xi_i$ and $\delta \omega_i$ along the $e_3$ axis. So we multiply the vector $\omega$ and $\xi$ with the matrix $C^\intercal$ to omit the third component \cite{b1}.

It can be shown that \begin{equation}\label{deltaqdot2}
    \begin{aligned}
        \delta \dot{q}_i = \delta \omega_i \times e_3
    \end{aligned}
\end{equation} at $q_i = e_3$.
Thus the state vector of the linearized equation is given by,
\begin{equation*}\begin{aligned} X = \begin{bmatrix}
    \delta v \\
    C^\intercal \delta \omega_1 \\
    C^\intercal \delta \omega_2 \\
    \vdots\\
    C^\intercal \delta \omega_n\\
    \delta x\\
    C^\intercal  \xi_1\\
    C^\intercal  \xi_2\\
    \vdots\\
    C^\intercal  \xi_n
    \end{bmatrix}\end{aligned}\end{equation*}.

Let \begin{equation*}
    \begin{aligned}
        L_1 =  \begin{bmatrix}
    M_{xx} & M_{xq}\\
    M_{qx} & M_{qq}
    \end{bmatrix}
    \end{aligned}
\end{equation*}
where the blocks are defined as:
\begin{equation*}
    \begin{aligned}
    M_{xx} &= M_{00} I_2\\
    M_{xq} &= \begin{bmatrix}
    -s_1 M_{01} \hat{e}_3 C & . & . & . & -s_n M_{0n} \hat{e}_3 C
    \end{bmatrix}\\
    M_{qx} &= M_{xq}^T\\
    M_{qq} &= \begin{bmatrix}
    M_{11} I_2 & s_{12} M_{12} I_2 & . & . & . & s_{2n} M_{1n} I_2\\
    s_{21} M_{21} I_2 & M_{22} I_2 & . & . & . & s_{2n} M_{2n} I_2 \\
    . \\ 
    . \\
    . \\
    s_{n1} M_{n1} I_2 & s_{n2} M_{n2} I_2 & . & . & . & M_{nn} I_2
    \end{bmatrix}\\
    G_{qq} &= \mbox{diag}[s_1 \sum_ {a = 1}^n m_a g l_1 I_2, .... \hspace{0.1cm},s_n m_n g l_n I_2 ]
    \end{aligned}
    \end{equation*}
The following equations give the linearized equations of motion:
\begin{equation}\label{linearized_equations}
    \begin{aligned}
        \left[ 
\begin{array}{c|c} 
  L_1 & 0_{2n+2} \\ 
  \hline 
  0_{2n+2} & I_{2n+2} 
\end{array} 
\right]  \dot{X} + &\begin{bmatrix}
   0_2 & 0_{2 \times 2n}& 0_2 & 0_{2 \times 2n}\\
   0_{2n \times 2} & 0_{2n} & 0_{2n \times 2} & G_{qq}\\
   I_{2} & 0_{2 \times 2n} & 0_2 & 0_{2 \times 2n}\\
   0_{2n \times 2} & I_{2n} & 0_{2n \times 2} & 0_{2n}
    \end{bmatrix} X = \\ &\begin{bmatrix}
    I_2\\
    0_{(4n+2) \times 2}
    \end{bmatrix}w + \begin{bmatrix}
    0_{2 \times 2}\\
    I_2 \\
    0_{4n \times 2}
    \end{bmatrix}u\\
    y &= \begin{bmatrix}0_{2 \times 2n} & I_2& 0_{(2n+2) \times 2}\end{bmatrix}X
    \end{aligned}
\end{equation}

Hence we see here the $A$, $B$, $E$ and $H$ matrices are given by:
\begin{equation}\label{ABEH_matrices}
    \begin{aligned}
        A &=  -\left[ 
\begin{array}{c|c} 
 L_1 & 0_{2n+2} \\ 
  \hline 
  0_{2n+2} & I_{2n+2} 
\end{array} 
\right]^{-1}\\
& \hspace{1cm} \begin{bmatrix}
   0_2 & 0_{2 \times 2n}& 0_2 & 0_{2 \times 2n}\\
   0_{2n \times 2} & 0_{2n} & 0_{2n \times 2} & G_{qq}\\
   I_{2} & 0_{2 \times 2n} & 0_2 & 0_{2 \times 2n}\\
   0_{2n \times 2} & I_{2n} & 0_{2n \times 2} & 0_{2n}
    \end{bmatrix}\\
    B &= \left[ 
\begin{array}{c|c} 
  L_1 & 0_{2n+2} \\ 
  \hline 
  0_{2n+2} & I_{2n+2} 
\end{array} 
\right]^{-1}\begin{bmatrix}
    0_{2 \times 2}\\
    I_2 \\
    0_{4n \times 2}
    \end{bmatrix}\\
    E &=  \left[ 
\begin{array}{c|c} 
  L_1 & 0_{2n+2} \\ 
  \hline 
  0_{2n+2} & I_{2n+2} 
\end{array} 
\right]^{-1}\begin{bmatrix}
    I_2\\
    0_{(4n+2) \times 2}
    \end{bmatrix}\\
    H &= \begin{bmatrix}0_{2 \times 2n} & I_2& 0_{(2n+2) \times 2}\end{bmatrix}
    \end{aligned}
\end{equation}
The output matrix is chosen such that the angular velocity of the $n^{th}$ link is the output of the system. This output is then decoupled from the disturbance input and then is integrated according to equation (\ref{kinematic_eq}) to give the direction vector of the $n^{th}$ link. Thus the direction vector of the $n^{th}$ link is decoupled from the disturbance input coming at the cart.
\section{Disturbance decoupling applied to the system}
In this section, we prove that it is possible to decouple the position or angular velocity of the $n^{th}$-link of an $n$-link chain pendulum on a cart from the disturbance coming at the cart with the help of a suitable state feedback. This result is proposed in the form of a theorem as follows.

\begin{Theorem}\label{main_theorem}
It is possible to design a state feedback $u=Fx$ for an $n$-link pendulum on a cart, such that the position or angular velocity of the last link can be decoupled from the disturbance input coming at the cart, provided $n \geq 2$.
\end{Theorem}
\begin{proof}
For this proof, we use the disturbance decoupling condition stated in (\ref{dist_decoup_cond}) and the properties of controlled invariant subspaces as stated in Theorem \ref{properties}.

It can be verified that the basis of kernel of the $H$ matrix in equation (\ref{ABEH_matrices}) does not have components in $(2n+1)^{th}$ and $(2n+2)^{th}$ axes. It can also be shown that the $B$ and $E$ matrices have the following structure:
\begin{equation*}
    \begin{aligned}
        B &= \begin{bmatrix}
            C^\intercal \hat{e}_3C\frac{1}{ml_1}\\
            \frac{m+m_1}{mm_1l_1^2}I_2\\
            -\frac{1}{m_1l_1l_2}I_2\\
            0_2\\
            \vdots\\
            0_2 \end{bmatrix}= \begin{bmatrix}
                *\\
                *\\
                *\\
                0_2\\
                \vdots\\
                0_2
            \end{bmatrix}
        \\
        E &= \begin{bmatrix}
            \frac{1}{m} I_2\\
            -C^\intercal \hat{e}_3C \frac{1}{ml_1}\\
            0_2\\
            \vdots\\
            0_2
        \end{bmatrix}
        =\begin{bmatrix}
            *\\
            *\\
            0_2\\
            \vdots\\
            0_2
        \end{bmatrix}
    \end{aligned}
\end{equation*}
It can be seen that the columns of $E$ are in the kernel of $H$. We will construct $\V^*$ such that it contains $\im(E)$.

We will now use the following property of controlled invariant subspaces:
\begin{equation*}
    A\V \subset \V + \im(B)
\end{equation*}

 We will choose a vector $v \in \V^*$ such that $AE = v + BFx$. This will prove the aforementioned property of controlled invariant subspaces thus proving that the image of $E$ is in $\V^*$ and $\V^*$ is controlled invariant.

It can be verified that 
\begin{equation*}
    \begin{aligned}
        AE = \begin{bmatrix}
            0 & 0\\
            \vdots & \vdots\\
            1/m & 0\\
            0 & 1/m\\
            0 & \frac{1}{ml_1}\\
            -\frac{1}{ml_1}&0\\
            0 & 0\\
            \vdots & \vdots\\
            0 & 0
        \end{bmatrix}
    \end{aligned}
\end{equation*}
From the structure of the $AE$, $E$ and $B$ matrices as shown above and the structure of the $A$ matrix, it can be shown:
\begin{enumerate}
    \item The non-zero entries of the $AE$ matrix are not in the $(2n+1)^{th}$ and $(2n+2)^{th}$ rows.
    \item Since $E$ contains non-zero entries in the first 4 rows, in order to contain $\im(E)$, a vector $v \in \V^*$ can be chosen such that it has non-zero entries in the first 4 rows and non-zero entries in the other rows similar to the $AE$ matrix.
    \item Now $B$ contains non-zero entries upto the 6th row. 
    \item Thus an input $u = Fx$ can be chosen such that when it is multiplied with $B$, it cancels out the entries of $v$ in the first 6 rows.
\end{enumerate}
Thus it can be shown that $AE = v + BFx$ for a $v \in \V^*$, which implies $AE \subset \V^* + \im(B)$.

The $(2n+1)^{th}$ and $(2n+2)^{th}$ rows of matrix $A$ can be showed to have the following structure:
\begin{equation*}
\begin{aligned}
    A((2n+1):&(2n+2),:) \\=& \begin{bmatrix}
        0_{4n \times 2} & \frac{\sum_{i = n-1}^{n}m_ig}{m_{n-1}l_n}I_2 & -\frac{\sum_{i = n-1}^{n}m_ig}{m_{n-1}l_n}I_2
    \end{bmatrix} \\=& \begin{bmatrix}
        0_{4n \times 2} & * & -*
    \end{bmatrix}
\end{aligned}
\end{equation*}
$V^*$ can be chosen as the set of basis vectors of subspace of kernel of $H$ which has equal components in $(4n+1)^{th}$ and $(4n+3)^{th}$ directions as well as $(4n+2)^{th}$ and $(4n+4)^{th}$ directions. This ensures the invariance condition and that $\V^*$ is in the kernel of $H$.

Thus it is possible to find an $(A+BF)$- invariant subspace such that it contains image of $E$ and is contained in the kernel of $H$.

This satisfies the disturbance decoupling condition and hence completes the proof.
\end{proof}

Thus it is possible to design a state feedback which decouples the output of a $n$-link pendulum from the disturbance coming at its base. We look at a numerical example of this in the next section.
\section{Numerical Example}
A 4-link pendulum on a cart in hanging configuration is selected. The specifications of this system are as follows:
\begin{enumerate}
    \item mass of the cart, $m_c = 4$
    \item masses of the links, $m = \begin{bmatrix}
    6 & 4 & 3 & 2
    \end{bmatrix}$
    \item lengths of the links, $l =\begin{bmatrix}
        5 & 4 & 2 & 2
    \end{bmatrix}$
    \item $s = \begin{bmatrix}
        1 & 1 & 1 & 1
    \end{bmatrix}$
    \item initial angle $\theta_i = \begin{bmatrix}
        89 \degree & 90\degree & 1\degree
    \end{bmatrix}$
\end{enumerate}

\begin{figure}[!htb]
\centering
\begin{subfigure}{.5\textwidth}
  \centering
  \includegraphics[width=.95\linewidth]{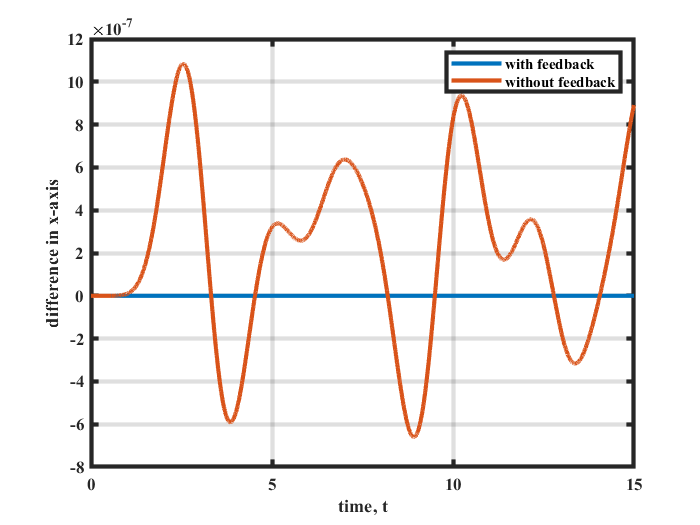}
  \caption{difference in x-axis}
  \label{fig:sfig1}
\end{subfigure}\\
\begin{subfigure}{.5\textwidth}
  \centering
  \includegraphics[width=.95\linewidth]{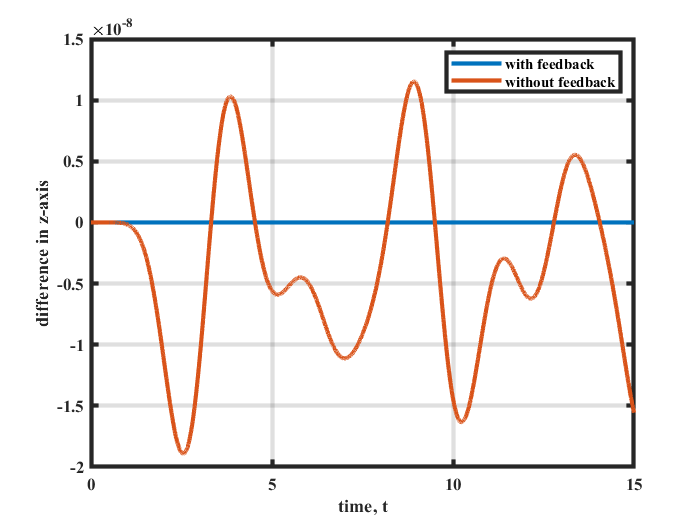}
  \caption{difference in z-axis}
  \label{fig:sfig2}
\end{subfigure}
\caption{Effect of disturbance decoupling}
\label{fig:effectofddp}
\end{figure}

The feedback matrix for disturbance decoupling for this system is computed using MATLAB and the algorithms stated in \cite{b2},\cite{b3}. The feedback matrix comes out to be 
\begin{equation*}
    \begin{aligned}
        F = \begin{bmatrix}
            0_{2 \times 12} & -735.75I_2 & 0_{2 \times 3} & 735.75D_2 & 0_{2 \times 1}
        \end{bmatrix}
    \end{aligned}
\end{equation*}
where $D_2 = \begin{bmatrix}
    0 & 1\\1& 0
\end{bmatrix}$.
Fig \ref{fig:effectofddp} shows the difference in output trajectory with and without feedback when 2 different disturbance inputs are applied to this system in the $x$-axis.

\begin{Note}
The orange and blue curves are differences in the trajectory of the $n^{th}$-link of the system in the $x$- and $z$- axes with and without feedback. They are not the actual trajectories.
\end{Note}
As can be seen in the figure \ref{fig:effectofddp}, the orange curve is the difference in output trajectories when there is no feedback in the system and the blue line is the difference in output trajectories when the feedback is introduced in the system. Fig \ref{fig:effectofddp} makes it clear that the feedback matrix computed completely decouples the output from the disturbance input.

The experiment was repeated for $n \in [2,8]$. It was found that the output of the n-link pendulum on a cart system can be decoupled from the disturbance input provided $n \geq 2$. This holds true for the inverted, hanging and folded equilibria for these configurations. Thus this experiment supports Theorem \ref{main_theorem}, which we have proposed and proved in the previous section.
\section{Conclusion and Future Works} 
In this paper, we solved Problem \ref{Problem2} where we applied the problem of disturbance decoupling to an $n$-link pendulum on a cart and showed that, by designing a suitable state feedback, it is possible to decouple the position or the angular velocity of the $n^{th}$-link of this system from the disturbance input.

 Future work in this area will focus on the position of the actuators in the system. In other words, it means at which link, the control needs to be applied to achieve perfect disturbance decoupling or disturbance decoupling with stability. It will also focus on developing the disturbance decoupling algorithm for the non-linear system itself, without linearizing it. In this context, the literature explored in \cite{b6} can be useful. In this paper, the procedure for computing a suitable feedback matrix for a non-linear system has been studied. Much like the case for a linear system, this involves constructing a maximal controlled invariant distribution contained inside kernel of the output matrix and contains the image of the disturbance input matrix. The concept of controlled invariance for non-linear systems with disturbances and several theorems have been discussed in this paper \cite{b6}. However, an algorithm suitable for computing the controlled invariant distribution on the manifold of the system needs to be designed. 
 
\end{document}